\documentclass[procedia]{easychair}

\usepackage{doc}
\usepackage{makeidx}
\usepackage{bm}

\newcommand{\mib}[1]{\bm{#1}}

\def\procediaConference{99th Conference on Topics of
  Superb Significance (COOL 2014)}

\title{Quantum Interference of Surface-Induced Friedel Oscillations Enhanced by Fermi-Surface Nesting in Layered Manganites}

\titlerunning{Valence Imbalance of Manganese Ions \ldots}

\author{Ryosuke Yamamura and Takashi Hotta}

\institute{Tokyo Metropolitan University, Hachioji, Tokyo, Japan　\\
              \email{yamamura-ryousuke@ed.tmu.ac.jp} \\
              \email{hotta@phys.se.tmu.ac.jp}}

\authorrunning{Yamamura and Hotta}

\begin{document}

\maketitle


\begin{abstract}
Imbalance of Mn$^{3+}$ and Mn$^{4+}$ ions between surface and bulk in layered manganites
is investigated on the basis of an orbital degenerate double-exchange model with surfaces.
For two types of $t_{\rm 2g}$ spin structures, the number of Mn$^{4+}$ ions on the surface is
found to be larger than that in the bulk, explained by the surface-induced Friedel oscillation.
In particular, for the sheet-type antiferromagnetic state, quantum beat phenomenon appears
due to the interference among Friedel oscillations and the number of Mn$^{4+}$ ions
tends to increase on the surface.
The present results are believed to be useful to develop high-efficient cathodes
in Li-ion batteries and catalysts.
\end{abstract}


%
%


\section{Introduction}
\label{sect:introduction}

In recent decades, the investigation of remarkable colossal magneto-resistance
(CMR) phenomena in manganites has been one of important research areas
in condensed matter physics \cite{Tokura}.
In one word, the CMR effect occurs when the ground state changes
from insulating to ferromagnetic (FM) metallic phases by the application of
a small magnetic field \cite{Dagotto1,Dagotto2,Hotta1,Hotta2}.
Among the two phases, the appearance of the FM metallic phase
has been usually rationalized by the so-called double-exchange mechanism
\cite{Zener} on the basis of a strong Hund coupling between
itinerant $e_{\rm g}$ electrons and localized $t_{\rm 2g}$ spins.
On the other hand, the insulating phase is basically
understood by the coupling between degenerate $e_{\rm g}$ electrons
and Jahn-Teller distortions of the MnO$_6$ octahedra,
leading to the various types of charge and/or orbital orders \cite{Dagotto1,Dagotto2,Hotta1,Hotta2}.

From the viewpoint of industrial application, manganites have been focused
as important materials for the magnetic head of high-density hard disk \cite{Grunberg}.
Furthermore, some manganites have been used as cathodes in Li-ion batteries
and catalysts.
In fact, as for the cathodes, LiMn$_{2}$O$_{4}$ with spinel structure has been used
for electric vehicles \cite{LiMn2O4} and, LiMnO$_{2}$ and their other transition metals-substituted materials 
with layered structure are expected to be promising cathode materials \cite{LiMnO2, LiNiMnO2}.
On the other hand, concerning the catalysts, much attention has been paid to
CaMn$_{4}$O$_{5}$ cluster in photosystem \cite{CaMn4O5}.
In artificial photosynthesis, MnO$_{2}$ has been investigated as a catalyst \cite{MnO2a,MnO2b},
instead of CaMn$_{4}$O$_{5}$ cluster from the viewpoint of the difficulty in synthesis.

In previous research on manganites,
the spin-charge-orbital ordering in the bulk
\cite{Dagotto1,Dagotto2,Hotta1,Hotta2}
and the interface in the heterostructures \cite{Yu}
have been discussed.
However, for the application to cathodes and catalysts,
it is important to consider explicitly {\it surface} electronic states.
It is well known that manganites work as cathodes and catalysts
when the valence changes mainly from Mn$^{4+}$ to Mn$^{3+}$
in the process of chemical reaction.
Thus, we focus on the valence change from Mn$^{4+}$ to Mn$^{3+}$ here.
To achieve high efficiency as cathodes and catalysts,
it is desirable to obtain the situation with Mn$^{4+}$ ions on the surface
and Mn$^{3+}$ in the bulk.
Although such a situation is naively expected to occur in metallic materials,
it is still unclear what microscopic mechanisms cause such valence imbalance
of manganese ions between surface and bulk.

In this paper, we investigate the effect of surface on the electronic state
in layered manganites on the basis of
an orbital-degenerate double-exchange model with surfaces.
For two kinds of $t_{\rm 2g}$ spin structures,
we find that the number of Mn$^{4+}$ ions on the surface is always
larger than that in the bulk, which is understood by
the surface-induced Friedel oscillation.
In particular, for the sheet-like antiferromagnetic (AF) phase,
we observe a peculiar beat phenomenon, explained by
the interference among Friedel oscillations.
In this case, the number of Mn$^{4+}$ ions more increases on the surface. 
We believe that the present results contribute to the development of
high-efficient cathodes and catalysts.
Throughout this paper, we use such units as $k_{\rm B}$=$\hbar$=$1$.

\section{Model and Method}
\label{sect:model and method}

To approach the present problem,
we introduce an orbital degenerate double-exchange model.
To simplify the model,
we adopt the widely-used approximation
with infinite Hund coupling between $e_{\rm g}$ and
$t_{\rm 2g}$ electron spins.
In such a limit, the $e_{\rm g}$ electron spin perfectly aligns along
the $t_{\rm 2g}$ spin direction,
leading to the suppression of spin degrees of freedom of $e_{\rm g}$ electrons.
Namely, $e_{\rm g}$ electrons move only in the FM region of $t_{\rm 2g}$ spins,
while in the AF region, $e_{\rm g}$ electrons are localized.
The Hamiltonian is given by
\begin{equation}
 H =  -\sum_{\mib{i} \mib{a} \gamma \gamma'}
        D_{\mib{i},\mib{i}+\mib{a}} t^{\mib{a}}_{\gamma \gamma'}
        d^\dagger_{\mib{i}\gamma}d_{\mib{i}+\mib{a}\gamma'} 
     + J \sum_{\langle \mib{i},\mib{j} \rangle} S_{z \mib{i}} S_{z \mib{j}},
\end{equation}
where $d_{\mib{i} a}$ ($d_{\mib{i} b}$) is the annihilation operator
for a spinless $e_{\rm g}$ electron in the $d_{x^{2}-y^{2}}$
($d_{3z^{2}-r^{2}}$) orbitals at site $\mib{i}$,
$\mib{a}$ is the vector connecting nearest-neighbor sites,
$D_{\mib{i},\mib{i}+\mib{a}}=(1+S_{z \mib{i}}S_{z \mib{i}+\mib{a}})/2$,
$S_{z\mib{i}}$ is Ising-like $t_{\rm 2g}$ spin at site $\mib{i}$
with $S_{z\mib{i}}=\pm 1$,
and $t^{\mib{a}}_{ \gamma \gamma'}$ denotes the nearest-neighbor
hopping amplitude between $\gamma$ and $\gamma'$ orbitals
along the $\mib{a}$ direction with
$t^{\mib{x}}_{aa}=-\sqrt{3}t^{\mib{x}}_{ab}
=-\sqrt{3}t^{\mib{x}}_{ba}=3t^{\mib{x}}_{bb}=3t^{\mib{z}}_{bb}/4$,
$t^{\mib{y}}_{aa}=\sqrt{3}t^{\mib{y}}_{ab}
=\sqrt{3}t^{\mib{y}}_{ba}=3t^{\mib{y}}_{bb}=3t^{\mib{z}}_{bb}/4$,
and $t^{\mib{z}}_{aa}=t^{\mib{z}}_{ab}=t^{\mib{z}}_{ba}=0$.
Hereafter we set the energy unit as $t^{\mib{z}}_{bb}=1$.
In the second term, $J$ is the AF coupling between
nearest-neighbor $t_{\rm 2g}$ spins and
$\langle \mib{i},\mib{j} \rangle$ denotes the nearest-neighbor site pair.
In this paper, we do not include explicitly the Coulomb interaction terms,
but the effect is partially considered by working with
a large Hund coupling.
Namely, we consider the manganite with large bandwidth.
We define $n$ as $e_{\rm g}$ electron number per site,
while $x$ is the hole doping with the relation of $x=1-n$.
Thus, $x$ indicates the average number of Mn$^{4+}$ ions per site.

In this work, the Hamiltonian is diagonalized
in a $1000 \times 1000 \times L$ lattice:
One layer is composed of $1000 \times 1000$ square lattice
with periodic boundary condition (PBC) along the x- and y-axes.
To introduce the surfaces in the model, we consider the stack of
$L$ layers with open boundary condition (OBC) along the z-axis.
%
In this paper, we introduce the OBC in the sense of $\psi (0)=\psi (L+1)=0$,
where $\psi(z)$ denotes the wavefunction of spinless $e_{\rm g}$ electron
along the z-axis and $z$ is a layer number.
%
Namely, the first layer along the z-axis indicates the surface,
whereas the $L/2$-th layer denotes the bulk.
In most cases, the cluster of $L=100$ is used,
while in a certain case, we consider the cluster of $L=1000$.


\begin{figure}
\centering
\includegraphics[width=\columnwidth,keepaspectratio]{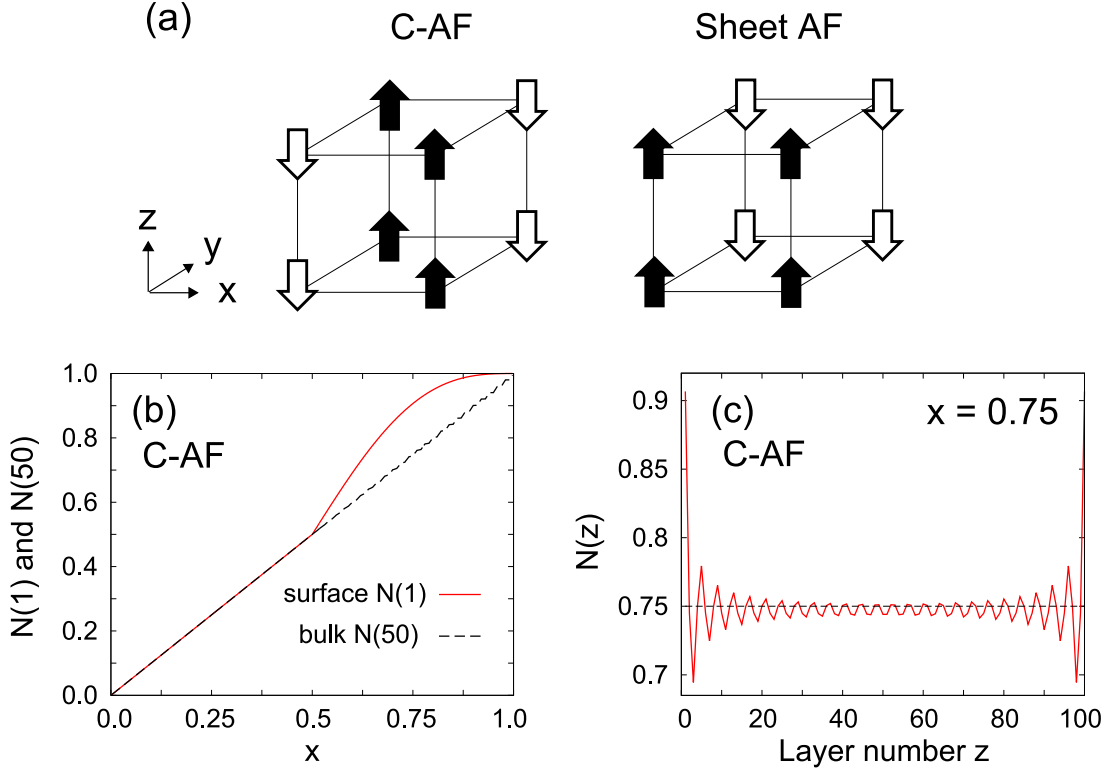}
\caption{
(a) Schematic views of spin structures used in this paper.
Solid and open arrows indicate up and down $t_{\rm 2g}$ spins, respectively.
To save space, we show the spin structure only
in a $2 \times 2 \times 2$ lattice.
(b) $N(1)$  (red solid curve)  and $N(50)$ (black dashed line)
as functions of hole doping $x$
for the $1000 \times 1000 \times 100$ lattice in the C-AF state.
(c) $N(z)$ vs. $z$ for $x=0.75$.
Broken horizontal line denotes the value of $x$.
}
\label{fig:fig1ICM}
\end{figure}

To concentrate on the effect of surface,
we do not discuss the stability of the ground-state magnetic
structures in this paper.
Rather we assume two kinds of spin structures:
C-type AF (C-AF) state and A-type AF (A-AF) state, as shown in Fig.~1(a).
At least in the $2 \times 2 \times 2$ lattice,
it is easy to show that such magnetic states actually
appear as the ground states for appropriate values of $J$.
In the C-AF state, $e_{\rm g}$ electron can move only
along the z-axis because of the infinite Hund coupling.
Thus, the system is regarded as a one-dimensional (1D) chain
with the OBC along the z-axis.
On the other hand, in the A-AF state,
$e_{\rm g}$ electron can move in the x-z plane,
leading to a two-dimensional (2D) sheet with edges.
Thus,  this spin structure is called the sheet AF state.
Due to the limitation of the page numbers,
we do not show in this paper
the calculation results for the three-dimensional
ferromagnetic state (3D FM),
where $e_{\rm g}$ electron can move in all three dimensions.

By using these lattices, we evaluate $N(z)$,
the number of Mn$^{4+}$ ions per site of the layer $z$,
to clarify the effect of the dimension of the FM region
in which $e_{\rm g}$ electrons can move.
We note that $N(1)$ and $N(L/2)$ denote the number of
Mn$^{4+}$ ions per site on the surface and in the bulk, respectively.


\section{Calculation Results}
\label{sect:calculation  results}


\subsection{C-type AF state}
\label{sect:c-type AF state}

First let us consider the charge structure in the C-AF state
for $L=100$.
In Fig.~1(b), we show $N(1)$ and $N(50)$ as functions of $x$.
For $0.0 \leq x \leq0.5$, $N(1)$ is exactly equal to $N(50)$.
In 1D chain along the z-axis, $d_{3z^{2}-r^{2}}$ orbital has itinerant nature,
while $d_{x^{2}-y^{2}}$ orbital becomes perfectly localized.
For $0.5 \leq x \leq 1$, $e_{\rm g}$ electrons exclusively accommodate
the band composed of $d_{3z^{2}-r^{2}}$ orbitals.
However, for $x \leq 0.5$, $e_{\rm g}$ electrons begin to accommodate
the flat band composed of localized $d_{x^{2}-y^{2}}$ orbitals.
Thus, the difference in charge densities between surface and bulk does not
appear in the region of $0.0 \leq x \leq 0.5$.
For $0.5 \leq x \leq 1.0$, $N(1)$ is clearly larger than $N(50)$.
Since $e_{\rm g}$ electrons cannot gain enough kinetic energy along the z-axis
at the edge, $e_{\rm g}$ electron number should be reduced on the surface,
leading to the increase of $N(1)$.
Namely, it is interpreted that a strong impurity potential exists at the edge.
In such a situation, we expect the emergence of surface-induced Friedel oscillation \cite{Friedel,Shibata}.

To confirm this expectation, we plot $N(z)$ vs. $z$ for $x=0.75$ in Fig.~1(c).
Note that $N(1)$ is larger than $x$.
We observe that $N(z)$ exhibits a damped oscillation from the surface ($z=1$)
to the bulk ($z=50$), which is well fitted by the formula for Friedel oscillation
as \cite{Friedel,Shibata,FO1,FO2,FO3}
\begin{equation}
  N(z)-x \propto \cos({2k_{\rm F}z}) z^{(-1-K_{\rho})/2},
\end{equation}
where $k_{\rm F}$ is the Fermi momentum
and $K_{\rho}$ is the correlation exponent \cite{Schulz}.
For $x=0.75$, $k_{\rm F}$ is given by $k_{\rm F}=\pi/4$,
consistent with the period of the oscillation in Fig.~1(c).
We note that $K_{\rho}=1$ in the present case,
since we do not include Coulomb interactions among $e_{\rm g}$ electrons.
In fact, the envelope of the oscillation is expressed by $1/z$.
Note also that $N(50)$ is slightly smaller than $x$.
The Friedel oscillation occurs due to the optimal $e_{\rm g}$ electron arrangement
so as to maximize the energy gain of kinetic motion.
However, the localized orbital does not contribute to the kinetic energy and
thus, we do not find Friedel oscillation in the region of $0.0 \leq x \leq 0.5$.
This is characteristic to the 1D $e_{\rm g}$-electron system.


\begin{figure}[t]
\centering
\includegraphics[width=\columnwidth,keepaspectratio]{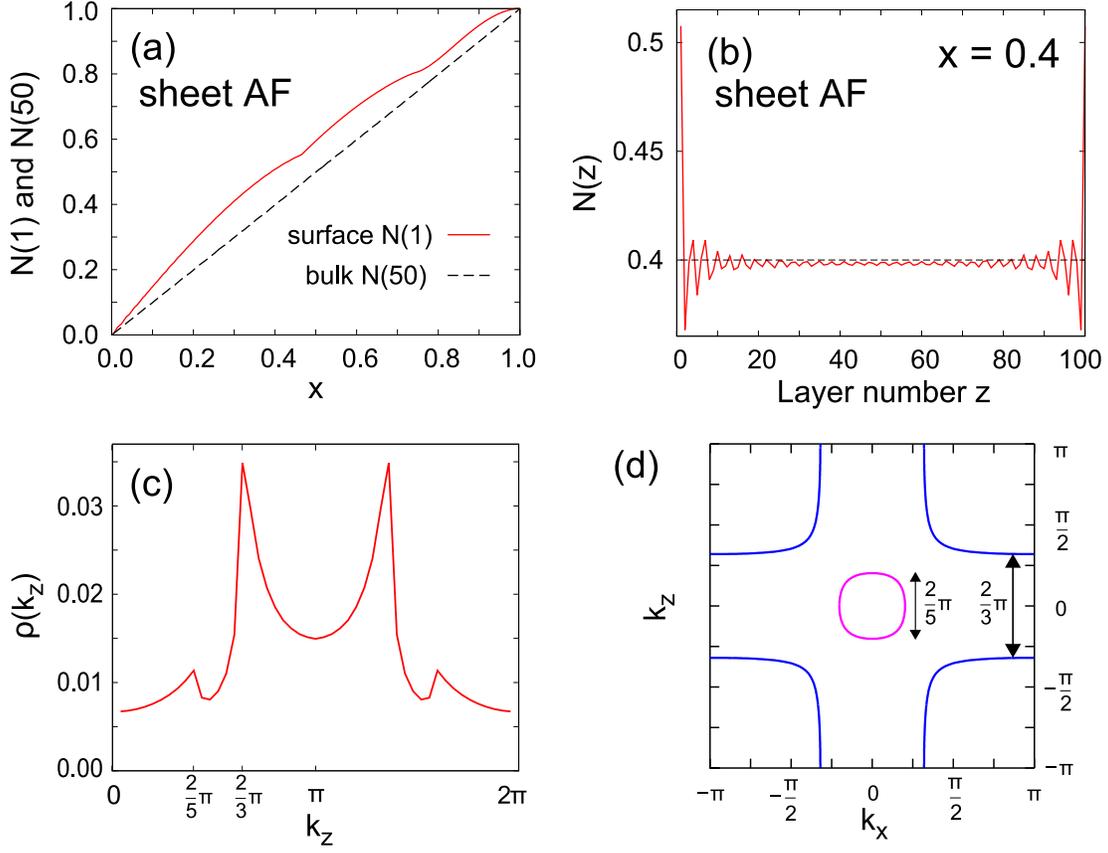}
\caption{
(a) $N(1)$ (red solid curve) and $N(50)$ (black dashed line)
as functions of $x$ for the $1000 \times 1000 \times 100$ lattice
in the sheet AF state.
(b) $N(z)$ vs. $z$ for $x=0.4$.
Broken horizontal line denotes the value of $x$.
(c) Fourier transform $\rho(k_z)$ of $N(z)$.
There are two peaks at $k_z=2\pi/5$ and $2\pi/3$
in the region of $0 \leq k_z \leq \pi$.
(d) Fermi-surface curves of $e_{\rm g}$ electron systems for $x=0.4$
by assuming the PBC along the z-axis.
The magnitudes of nesting vectors along the z-axis are $2\pi/5$ and $2\pi/3$,
which are consistent with the peak positions in (c).
}
\label{fig:fig2}
\end{figure}

\subsection{Sheet-type AF state}
\label{sect:sheet-type AF state}

Next we consider the results in the sheet AF state for $L=100$.
As shown in Fig.~2(a), $N(1)$ is larger than $N(50)$
for the whole range of $x$ in contrast to Fig.~1(b) for the C-AF state.
In Fig.~2(b), we plot $N(z)$ for $x=0.4$.
We again observe a Friedel-like oscillation, but it is not obvious
at this stage that it can be really considered as the Friedel oscillation.
In contrast to the C-AF state, the impurity potentials
in the sheet AF state are regarded to array on the edge line and
the oscillation is not simply expressed by eq.~(2).

Let us examine the oscillation from a viewpoint of the Fermi-surface structure.
We show the Fourier transform $\rho(k_z)$ of $N(z)$ and the Fermi-surface curves
at $x=0.4$ in Fig.~2(c) and 2(d), respectively.
Note that we depict the Fermi-surface curves by assuming the PBC along the z-axis,
since in this case, we can ignore the effect of the difference between OBC and PBC.
In Fig.~2(c), we find two peaks in the region of $0 \leq k_z \leq \pi$
at $k_z=2\pi/5$ and $2\pi/3$,
suggesting that the oscillation includes two wavevectors.
By analogy with the wavevector of the Friedel oscillation expressed in eq.~(2),
we deduce that such two wavevectors are closely related with
nesting vectors along the z-axis
in the 2D FM region of the sheet AF state.
In fact, the nesting vectors are $2\pi/5$ and $2\pi/3$ in this case,
as shown in Fig.~2(d).
In the sheet AF state, since the average charge density per layer
oscillates along the z-axis,
we conclude that Fig.~2(b) denotes the Friedel oscillation
in the 2D system.
Note that the envelop of the 2D Friedel oscillation decays
faster than $1/z$, but the discussion on this point is postponed
in a future paper.


\begin{figure}[t]
\centering
\includegraphics[width=\columnwidth,keepaspectratio]{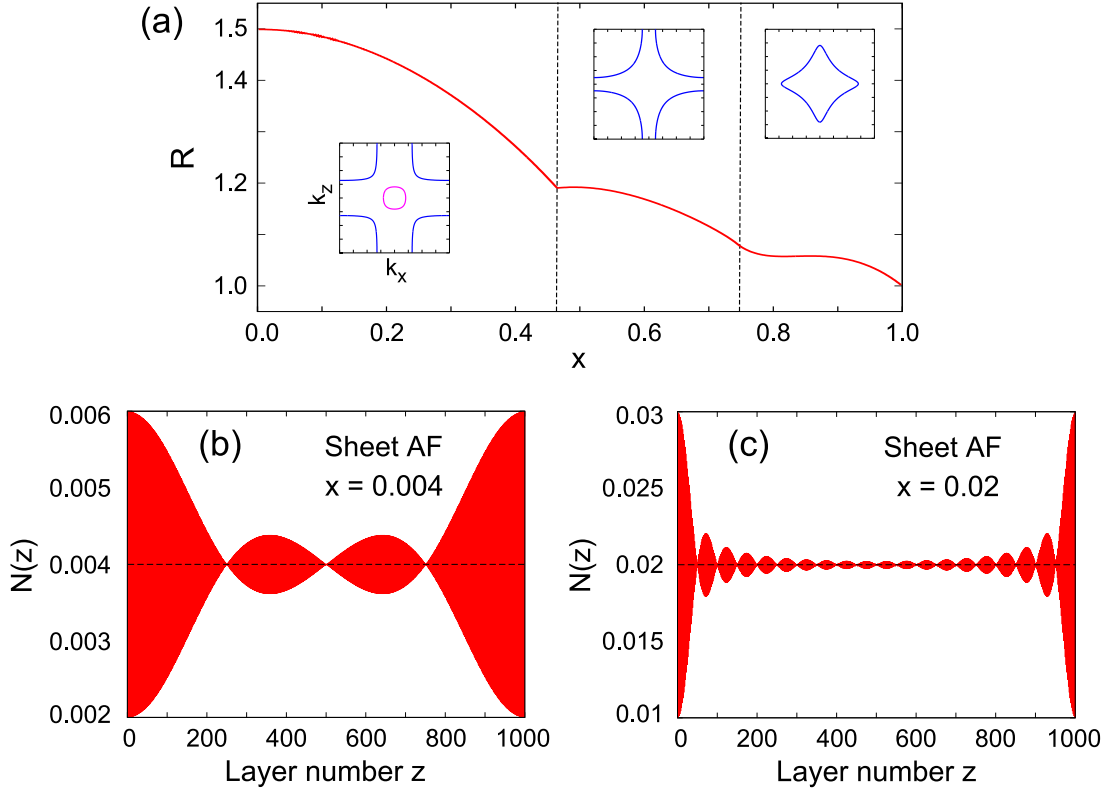}
\caption{
(a) Efficiency $R$ vs. $x$ for $1000 \times 1000 \times 1000$
lattice in the sheet AF state.
The insets denote typical Fermi-surface structures
which characterize three regions.
(b) $N(z)$ vs. $z$ for $x=0.004$.
Broken horizontal line denotes the value of $x$.
(c) $N(z)$ vs. $z$ for $x=0.02$.
This result can be well reproduced by the superposition of
ten oscillations.
}
\label{fig:fig3}
\end{figure}

\subsection{Efficiency and beat phenomenon for Sheet-type AF state}
\label{sect:efficiency and beat phenomenon for sheet-type AF state}

To clarify the increase in $N(1)$ in comparison with $x$,
we define the ratio $R$ as $R=N(1)/x$,
which expresses the degree of the appearance tendency of
Mn$^{4+}$ ions on the surface at a certain hole doping $x$.
When $N(1)$ is enhanced, we expect that the efficiency
as cathodes and catalysts is increased.
Thus, $R$ is called the ``efficiency'' in this paper.

For further investigation of physical properties
in the sheet AF state, we consider the cluster of $L=1000$.
In Fig.~3(a), we plot the efficiency $R$ as a function of $x$
for the sheet AF state.
We find three regions classified by the topology of
the Fermi-surface structure, shown in the insets.
In the region of $0 < x < 0.47$, we find two Fermi-surface curves
with the centers at $\Gamma$ and $M$ points.
In the regions of $0.47 < x < 0.75$ and $x > 0.75$,
we find one Fermi-surface curve with the center
at $M$ and $\Gamma$ points, respectively.
At the boundaries of these regions, we find a kink or an inflection point,
which are the so-called van Hove singularities.

Now we turn our attention to the small $x$ region,
where the magnitude of $R$ is enhanced.
In Fig.~3(b), we plot $N(z)$ at $x=0.004$,
in which we find a beat pattern in sharp contrast to
the damped oscillation observed in Fig.~2(b).
As discussed in Fig.~2(b), when two Fermi-surface curves exist,
$N(z)$ is considered to be given by the superposition of
two Friedel oscillations with wavevectors
corresponding to the nesting vectors along the z-axis.
In the 2D FM region, the spinless $e_{\rm g}$ electron system possesses
the Fermi-surface structure near the perfect nesting in the vicinity of
$x=0$ and such two nesting vectors along the z-axis take the similar values.
Thus, at a first glance, the interference between two Friedel oscillations
with almost the same wavevectors seems to induce a beat.

However, when we carefully investigate the wavevector of the beat,
we notice that it is not simply given by the difference
in two nesting vectors along the z-axis near $x=0$.
Rather the resolution $\delta k_z$ in the $\mib{k}$ space
along the z-axis plays a key role for the formation of the beat
in $N(z)$ near $x=0$.
When we consider $L$ layers along the z-axis in the OBC,
$k_z$ is given by $k_z(\ell)=\ell\pi/(L+1)$ with an integer $\ell$
($=1, 2, \cdots, L$), leading to $\delta k_z=\pi/(L+1)$.
%
This relation is obtained from the condition of $\psi (0)=\psi (L+1)=0$.
%
Then, we find a simple expression for $N(z)$ near $x=0$ as
\begin{equation}
  \label{eq3}
  N(z)=\sum_{\ell} A_{\ell} \sin ^2 [z k_{z}(\ell)],
\end{equation}
where $A_{\ell}$ is the amplitude for the oscillation with $k_z(\ell)$.
For small $x$, the deviation of the shape of the Fermi-surface curves
from the perfect nesting structure occurs in the unit of $\delta k_z$
and several numbers of oscillations with $k_z(\ell)$ near
$\ell=L/2$ (half-filling) contribute to $N(z)$ in eq.~(\ref{eq3}).
Thus, the difference between two nesting vectors does
not play an important role in $N(z)$ near $x=0$.
In fact, the pattern in Fig.~3(b) is well reproduced
by the superposition of two oscillations of $k_z(499)$ and $k_z(500)$
with the same amplitudes.

When we increase $x$,
the number of nodes in $N(z)$ is increased, 
even though two Fermi-surface curves continue to exist.
For instance, we plot $N(z)$ at $x=0.02$ in Fig.~3(c), which
cannot be explained by the superposition of two oscillations.
However, we can reproduce Fig.~3(c) by the superposition of
ten oscillations with $k_z(\ell)$ from $\ell=491$ to $\ell=500$
with the same amplitudes in eq.~(\ref{eq3}).
In this case, we find $19$ nodes in $N(z)$.
When we define $\ell_0$ as the number of superposed oscillations,
the number of nodes in $N(z)$ is given by $2\ell_0-1$
and the corresponding hole doping $x_0$ is expressed as $x_0=(2/L)\ell_0$.
For the value of $x$ deviated from $x_0$, there occurs a beat
corresponding to $x_0$ near $x$.

For $x > 0.2$, the resolution becomes relatively smaller
than the difference in two nesting vectors of the Fermi-surface curves.
In such a case, the beat phenomenon disappears and
instead the oscillation like Fig.~2(b) begins to appear.
The exotic beat phenomenon can be observed only in the small $x$ region,
where the effect of the resolution in the $\mib{k}$ space is significant.
We emphasize that the clear Friedel oscillation is observed
and $N(1)$ is enhanced only for the Fermi-surface structure
with good nesting properties.
In fact, when the nesting properties are lost for larger $x$,
we simply observe a monotonic decrease in $N(z)$ from surface to bulk.

\section{Discussion and Summary}
\label{sect:discussion and summary}

In this paper, we have discussed the dependence of charge densities
on the layer number in the orbital degenerate double-exchange model
with surfaces for C-AF and sheet AF states.
We have found that Mn$^{4+}$ number on the surface is
larger than that in the bulk, which is basically understood from
the surface-induced Friedel oscillations.

Concerning the future research directions on the present issue,
four comments are in order.
(i) In this paper, we have considered only two kinds of $t_{\rm 2g}$
spin patterns, C-AF and sheet AF states, mainly due to the limitation of
the page numbers.
For other $t_{\rm 2g}$ spin patterns such as 3D FM and zigzag sheet AF states,
we will show the results in a separate paper in near future.
(ii) Although we have simply assumed the $t_{\rm 2g}$ spin patterns,
it is highly desired to confirm the stability of those magnetic structures
in the large-sized cluster with surfaces.
A straightforward way is to perform the Monte Carlo simulations,
but such calculations will be done in near future.
(iii) Since we have considered the manganite with large bandwidth
in this paper,
we have simply ignored Coulomb interaction and Jahn-Teller distortions.
These effects will be significant in the manganite with narrow bandwidth.
%
%
In particular, when we introduce the Coulomb interaction
between $e_{\rm g}$ electrons,
we are interested in the effect of the scattering of $e_{\rm g}$
electrons on the Friedel oscillations.
Concerning the Jahn-Teller distortions,
we are also interested in the different Jahn-Teller distortions
on the surface and in the bulk, 
in addition to the effect of Jahn-Teller distortions on
the magnetic structure.
%
Note that in the $2 \times 2 \times 2$ and $4 \times 4 \times 4$ lattices,
we have confirmed the appearance of A-AF, C-AF and FM states
for appropriate values of $J$, when we include the
effect of Jahn-Teller distortions \cite{Hotta3,Hotta4}.
Such calculations with
%
%
Coulomb interaction and/or
%
Jahn-Teller distortions will be performed in future.
(iv) We have investigated static properties of surface electronic states,
but to consider the actual application of mangatites to cathodes
and catalysts, local dynamical properties on the surface are also important.
%
%
Furthermore, the direction of the surface and its condition may provide
significant effects on the Friedel oscillations emphasized in this paper.
These points are other future problems.
%

In summary, we have evaluated $N(z)$,
the number of Mn$^{4+}$ ions per site of the layer $z$,
by analyzing the orbital degenerate double-exchange model.
For the C-AF and the sheet AF states, we have found
$N(1) \geq N(50)$ due to the surface-induced Friedel oscillations.
In particular, in the sheet AF state, we have observed the beat phenomenon
due to the interference among Friedel oscillations,
leading to the increase of $N(1)$.
The present results provide fundamental understandings on further application
of manganites such as high-efficient cathodes in Li-ion batteries and catalysts.


\section*{Acknowledgment}

We thank K. Hattori, K. Kubo, and K. Ueda for discussions and comments.
The computation in this work has been partly done using the facilities of the
Supercomputer Center of Institute for Solid State Physics, University of Tokyo.


\end{document}